# Radiative heat exchange of spherical particles with metal and insulator plates

*G.V.Dedkov[1] and A.A.Kyasov*

Kabardino –Balkarian State University, Nalchik, 360004, Russian Federation

**Abstract**
Radiative heat exchange of spherical particles between each other and with thick polarizable plates is studied in the framework of fluctuation electrodynamics. An additive dipole approximation for the thermal conductance of micrometer sized particles is proposed. The numerical calculations are performed for particles and plates made of gold and silica. Several theoretical models of radiative conductance are compared and juxtaposed with the known experimental data.

**1.Introduction**

A theory of radiative heat exchange (RHE) between the bodies of different temperature divided by a narrow vacuum gap has been developed by several groups of authors [1-15]. Both RHE and conservative/dissipative van –der –Waals (Casimir) forces appear due to fluctuation electromagnetic interaction. A necessary condition for RHE is the temperature difference between the bodies, while the Casimir forces exist even between cold bodies.

However, analysis of literature shows that until now Casimir forces have attracted much more attention than RHE, both theoretically and experimentally (for a brief review see [15-17]). There are only a few geometrical configurations which have been studied in details: 1) two parallel plates [3,6-8]; 2) two spherical particles in the dipole approximation [11,12,19,20]; 3) a dipole particle above a flat polarizable surface (thick plate) [15]. Also, in Ref. [12], using the nonretarded approximation of fluctuation electrodynamics, the authors have studied RHE between multipolar particles.

---

[1] Corresponding author e-mail: gv_dedkov@mail.ru

In general, calculation of the structure of fluctuation electromagnetic field nearby arbitrary curved surfaces leads to severe technical problems, therefore above mentioned configurations have to be considered as the guiding ones.

Experimental investigations of RHE are also rare [22-28]. Only in [26-28] the authors have used configuration of a sphere on the flat, being more convenient under positioning of contacting bodies. In earlier papers [22-24] the configuration of two parallel plates has been used. The results obtained in Ref. [22] are thought to have only a qualitative nature [23]. In Ref.[23], on the other hand, a minimal approach distance between the plates has exceeded several $\mu m$, thus the radiation heat flux due to evanescent electromagnetic modes proved to be too small to be measured. The authors of [24] have used a configuration of the scanning probe microscope to measuring radiative conductance at a gap width in the range of $0.05 \div 0.2\,\mu m$. However, in our opinion, the interpretation of these results has some ambiguity due to the lack of information about an overall tip shape used. Very likely that the same puzzle is relevant to the results [25] obtained in the same configuration, where the radiation heat flux was measured at smaller gap widths ranging in the interval of $1 \div 100\,nm$.

The aim of this paper is twofold: 1) discussion and comparative numerical testing of our previously obtained theoretical expressions for RHE in several geometrical configurations; 2) comparison of the calculated radiative conductances with recently obtained experimental data [26-28].

**2. Theoretical outline**

General expression for the rate of radiative cooling (heating) of a small spherical particle with temperature $T_1$, placed near a plate with temperature $T_2$, both embedded into equilibrium photon gas with temperature $T_2$, has been obtained in our papers [15]. The corresponding formula is valid in the dipole approximation $R/z \ll 1, R \ll \lambda_0$

$$\dot{Q} = -\frac{4\hbar}{\pi c^3} \int_0^\infty d\omega \omega^4 [\alpha_e''(\omega) + \alpha_e''(\omega)][\Pi(\omega, T_1) - \Pi(\omega, T_2)] -$$

$$-\frac{2\hbar}{\pi c^3} \int_0^\infty d\omega \omega^4 [\Pi(\omega, T_1) - \Pi(\omega, T_2)] \int_0^\infty du \exp\left(-\frac{2\omega z}{c} u\right)(\alpha_e'' \operatorname{Im} f_e + \alpha_m'' \operatorname{Im} f_m) - \quad (1)$$

$$-\frac{2\hbar}{\pi c^3} \int_0^\infty d\omega \omega^4 [\Pi(\omega, T_1) - \Pi(\omega, T_2)] \int_0^1 du \left[ \operatorname{Re}\left(e^{\frac{2i\omega z}{c} u} \tilde{f}_e\right) \alpha_e'' + \operatorname{Re}\left(e^{\frac{2i\omega z}{c} u} \tilde{f}_m\right) \alpha_m'' \right]$$

$$f_e(u, \omega) = (2u^2 + 1)\Delta_e(u, \omega) + \Delta_m(u, \omega) \tag{2}$$

$$f_m(u, \omega) = (2u^2 + 1)\Delta_m(u, \omega) + \Delta_e(u, \omega) \tag{3}$$

$$\tilde{f}_e(u, \omega) = (1 - 2u^2)\tilde{\Delta}_e(u, \omega) + \tilde{\Delta}_m(u, \omega) \tag{4}$$

$$\tilde{f}_m(u, \omega) = (1 - 2u^2)\tilde{\Delta}_m(u, \omega) + \tilde{\Delta}_e(u, \omega) \tag{5}$$

$$\Delta_e(u, \omega) = \frac{\varepsilon(\omega)u - \sqrt{u^2 + 1 - \varepsilon(\omega)}}{\varepsilon(\omega)u + \sqrt{u^2 + 1 - \varepsilon(\omega)}}, \quad \Delta_m(u, \omega) = \frac{u - \sqrt{u^2 + 1 - \varepsilon(\omega)}}{u + \sqrt{u^2 + 1 - \varepsilon(\omega)}} \tag{6}$$

$$\tilde{\Delta}_e(u, \omega) = \frac{\varepsilon(\omega)u - \sqrt{u^2 + \varepsilon(\omega) - 1}}{\varepsilon(\omega)u + \sqrt{u^2 + \varepsilon(\omega) - 1}}, \quad \tilde{\Delta}_m = \frac{u - \sqrt{u^2 + \varepsilon(\omega) - 1}}{u + \sqrt{u^2 + \varepsilon(\omega) - 1}}$$

where $R$ is the particle radius, $z$ is the distance from the particle center to the plate and $\lambda_0$ is the characteristic wave length of absorption of electromagnetic radiation. In eq.(1), $\alpha_{e,m}'' \propto R^3$ denotes the frequency –dependent imaginary part of the particle electric and magnetic polarizability, $\varepsilon(\omega), \mu(\omega)$ denote dielectric and magnetic permittivities of materials, $\Pi(\omega, T) = 1/(\exp(\hbar\omega/k_B T) - 1)$ is the Bose –Einstein distribution for photons, other quantities have their usual meaning. For brevity, arguments of some functions in eq.(1) are omitted.

The first integral term in eq. (1) does not depend on $z$ and describes RHE between the particle and vacuum background. The second and third integral terms describe RHE between the particle and surface of the plate being mediated by near field and far field modes of fluctuation electromagnetic field.

For a neutral ground state atom, eq.(1) forecasts heating effect, which may be interpreted in terms of the shift of the atomic energy levels [15]. It should be noted that vacuum and surface contributions to the particle heating rate result from the lack of correlation between vacuum and surface modes of fluctuation electromagnetic field. As illustrated by detailed calculation, the second and third integrals in eq.(1) at large enough distances $z$ have an opposite sign and asymptotically cancel each other (see lines 1,2 on Fig.1). As a result, the particle heating (cooling) rate is determined by vacuum radiation only. Therefore, in this configuration, the corresponding RHE vacuum asymptotics is realized in a quite natural manner. On the other hand, if we do not take into account vacuum radiation (dashed –dotted line on Fig.1), this leads to a physically incorrect behavior of the heating rate at some distances $z$ (cf. dotted and dashed lines): $dQ/dt > 0$ at $T_1 < T_2$. This manifests principal role of vacuum background in configuration particle –plate [15].

It is worth noting that Stefan –Boltzmann's (SB) radiation law does not result from eq.(1) because in this case the condition $R >> \lambda_0$ must be fulfilled. In the SB limit, the radiation power has to be calculated from eq.(9) being multiplied by the particle surface area $4\pi R^2$. In the intermediate case $R \approx \lambda_0$ the radiation power must be determined using the theory developed by Mie [29].

Now, following [3,6,8,15], let us consider RHE in configuration of two thick parallel plates. An expression for the radiative heat flux between two plates (semi infinite homogeneous and isotropic spaces) with temperature $T_1$ and $T_2$, per unit surface area is given by

$$S(l) = \frac{\hbar}{4\pi^2} \int_0^\infty d\omega \omega [\Pi(\omega, T_1) - \Pi(\omega, T_2)] \cdot$$
$$\cdot \sum_{\mu=e,m} \left[ \int_0^{\omega/c} dk k \frac{(1 - |\Delta_{1\mu}|^2 - |\Delta_{2\mu}|^2)}{|D_\mu|^2} + 4 \int_{\omega/c}^\infty dk k \frac{\text{Im}\,\Delta_{1\mu}\,\text{Im}\,\Delta_{2\mu} \exp(-2q_0 l)}{|D_\mu|^2} + \right] \qquad (8)$$

where $q_0 = (k^2 - \omega^2/c^2)^{1/2}$, $D_\mu = 1 - \Delta_{1\mu}\Delta_{2\mu}\exp(-2q_0 l)$, subscripts 1,2 indicate the plates and $\mu -$ the waves of different polarization. The reflection amplitudes $\Delta_{i\mu}$ are given by (6),(7) for each plate and type of material and polarization. To express $\Delta_{i\mu}$ in terms of wave vector $k$, one should make substitution $u^2 + 1 = kc/\omega$ in eq.(6) and $1 - u^2 = kc/\omega$ in eq.(7).

The first term in eq.(8) weakly depends on distance $l$ between the plates and gives heat flux related with propagating modes. For black materials, $\Delta_{1\mu} = \Delta_{2\mu} = 0$, the corresponding heat flux is equal to a difference between the two heat fluxes coming from one plate to another –a classical SB result. The second term in eq.(8) corresponds to the heat flux related with evanescent modes of fluctuation field. Because in the involved problem statement an interaction of the plates with vacuum background is not taken into account, one can not obtain from (8) the heat flux radiated from a single isolated plate into the vacuum space [2]:

$$S^{vac} = \frac{\hbar}{4\pi^2} \int_0^\infty d\omega \omega [\Pi(\omega,T_1) - \Pi(\omega,T_2)] \int_0^{\omega/c} dk k (2 - |\Delta_{1e}|^2 - |\Delta_{1m}|^2) \tag{9}$$

Contrary to Eq.(9), even at $l \to \infty$ the first term of (8) manifests weak space oscillations. It must be noted that in eqs.(8), (9) $S > 0$ corresponds to the heat flux coming from a more heated plate to a colder one, whereas in equations for $dQ/dt$ like (1), (10), (14), the out coming heat flux will correspond to $dQ/dt < 0$.

Also, a closed analytical solution to the RHE problem has been obtained in configuration of two dipole particles. In the case when two spherical particles of radii $R_{1,2}$ and temperatures $T_{1,2}$ are embedded into equilibrium background radiation with temperature $T_3$, the resultant heating (cooling) rate of the first particle is given by [19,20]

$$dQ_1/dt = \frac{4\hbar}{\pi c^3} \int_0^\infty d\omega \omega^4 (\alpha''_{1e}(\omega) + \alpha''_{1m}(\omega))[\Pi(\omega,T_3) - \Pi(\omega,T_1)] +$$
$$+ \frac{4\hbar}{\pi r^6} \int_0^\infty d\omega\, \omega [\alpha''_{1e}(\omega)\alpha''_{2e}(\omega) + \alpha''_{1m}(\omega)\alpha''_{2m}(\omega)] \cdot (3 + (\omega r/c)^2 + (\omega r/c)^4) \cdot \tag{10}$$
$$\cdot [\Pi(\omega,T_2) - \Pi(\omega,T_1)]$$

where $r$ is the distance between the particle centers, $\alpha_{i\,e,m}(\omega)$ denotes polarizabilities of the different particles ($i = 1,2$) and type (subscripts "$e,m$"). The validity conditions for the used dipole approximation imply

$$R_1, R_2 \ll r, \quad R_1, R_2 \ll \min(\lambda_{T1}, \lambda_{T2}, \lambda_{T3}) \tag{11}$$

where $\lambda_{T1}, \lambda_{T2}, \lambda_{T3}$ are the characteristic wave lengths of the equilibrium thermal radiation with temperatures $T_1, T_2, T_3$.

It is worth noting that thermal configuration of two particles (eq.(10)) turns out to be even more general than configuration particle –plate, corresponding to eq.(1), where the plate is assumed to be in thermal equilibrium with surrounding background radiation. Similar to eq.(1), in eq.(10) the first integral term denotes the particle thermal radiation being emitted into (absorbed from) the vacuum space, while the second term denotes heat flux coming directly between the particle and plate.

In papers [19,20] the authors have calculated only the second term of eq.(10), depending on $r$. However, the numerical coefficient before the obtained integral expression turned out to be essentially different from $4/\pi$ in eq.(10). So, in Ref.[8] this coefficient was equal to $1/4\pi^2$, while in a more recent paper of these authors one finds $128/\pi$ [10]; $1/4\pi^3$ in Ref.[11] and $1/8\pi^3$ in Ref.[12]. The lack of concord in this problem may lead to a possible change in the calculated heating rate by $10^4$ times ! In our opinion, the involved disagreements are caused by numerical errors in definitions of correlators of the fluctuation electromagnetic fields and dipole moments.

To conclude this section, we note that consideration of the last geometrical configuration in view of goals raised in this work is motivated by two reasons: 1) it is of great importance to make a detailed comparison between the configurations sphere –plate and sphere –sphere; 2) to check an assumption [26] that in the former configuration the thermal conductance is twice that in the second configuration assuming the same value of gap width.

**3. Heating rate of a spherical particle in close vicinity to the plane surface**

For a big particle of radius $R$ being placed near a plate, eqs.(1), (8) are not valid. The same situation one faces when calculating the Casimir forces. Thus, for small gap widths, eq.(8) is being currently employed in an approximate way, making use the proximity force approximation (PFA) [17]. By analogy with that, to calculate radiative heat flux in the geometry sphere –plate, one should integrate eq.(8) over the contact area [8,10,28]

$$dQ/dt = -2\pi \int_0^\infty d\rho\, \rho\, S(z(\rho)) \approx -2\pi R \int_{z_0}^\infty S(z)dz \qquad (12)$$

where $z_0$ is the distance of minimal approach. A validity of eq.(12) is conditioned by $z_0/R \ll 1$. Particularly, in calculations of the Casimir forces, an error of PFA grows linearly with increasing $z_0/R$ [30].

In our recent papers [15,21] we have put forward another approach, based on using eq.(1) and additive dipole approximation (DAA). In DAA, eq.(1) is assumed to be a local relation for a small volume $dV$ of material of the particle. Then, substituting $R^3 \to (3/4\pi)dV$ in the polarizability coefficients, one has to integrate eq.(1) over the particle volume

$$dQ(z_0)/dt = \int_V f(z)d^3r = \pi z_0^3 \int_1^{1+2R/z_0} f(z_0 s)\left[\frac{2R}{z_0}(s-1) - (s-1)^2\right]ds, \quad s = z/z_0 \qquad (13)$$

where $f(z = z_0 s)$ is the integrand function, corresponding to the second and third terms in the right hand side of eq.(1). A contribution of the vacuum heating rate (the first term eq.(1)) in DAA will not change. A similar expression can be used to calculate the van –der -Waals and Casimir forces [15]. An advantage of eq.(13) is related with an absence of a restriction limit on $z_0/R$: at $z_0/R \gg 1$ eq.(13) becomes exact and reduces to (1). On the other hand, at $z_0/R \ll 1$ its accuracy is comparable with PFA, at least in calculations of the Casimir forces [15].

In addition, it has to be noted that DAA principally differs from the well known additive summation method used to calculating the van –der -Waals and Casimir forces between extended bodies via pair interatomic potentials [17]. Eq.(1) takes into account the retardation and temperature effects on the one hand, and non additive effects to some extent, on another hand, because the particle consists of many atoms.

To proceed further, it is convenient to change an order of integration over variables $u$ and $s$, when substituting (1) into (13), first making use integration over $s$ variable. The resulting expression for the surface contribution to the particle heating rate is given by

$$dQ_s/dt = -\frac{2\hbar}{\pi c^3}\int_0^\infty d\omega\, \omega^4 [\Pi(\omega,T_1) - \Pi(\omega,T_2)]\int_0^\infty du\, Y_1(xu,y)\cdot[\alpha_e'' \operatorname{Im} f_e + \alpha_m'' \operatorname{Im} f_m] -$$

$$-\frac{2\hbar}{\pi c^3}\int_0^\infty d\omega\, \omega^4 [\Pi(\omega,T_1) - \Pi(\omega,T_2)]\int_0^1 du\, \{[\operatorname{Re}\tilde{f}_e \cdot Y_2(xu,y) - \operatorname{Im}\tilde{f}_e \cdot Y_3(xu,y)]\alpha_e'' + (e \to m)\}, \qquad (14)$$

$x = 2\omega z/c$, $y = 2R/z_0$

where the term $(e \to m)$ is identical to the first one in the figure brackets, being taken with the proper change of the integrand function, the auxiliary functions $Y_{1-3}(x,y)$ are given in Appendix. Moreover, one should note that in eq.(14) the polarizability functions must be normalized per unit volume of particles.

Quite recently we have used DAA to calculate RHE between large spherical particles of radius $R$ being placed close to one another [19]. In this case, the second term of eq.(10) has to be integrated over the volume of the spheres and the resultant expression is given by (without of vacuum contribution)

$$dQ_{12}/dt = \frac{4\hbar}{\pi} \int_0^\infty d\omega [\alpha''_{1e}\alpha''_{2e} + \alpha''_{1e}\alpha''_{2e}][3f_1(x) + (\omega R/c)^2 f_2(x) + (\omega R/c)^4 f_3(x)] \cdot$$
$$\cdot [\Pi(\hbar\omega, T_2) - \Pi(\omega, T_1)], \quad x = 2R/(2R + z_0) \quad (15)$$

where $z_0$ is the gap width and the functions $f_{1-3}(x)$ are also given in Appendix. As in eq.(14), the polarizabilities in eq.(15) must be normalized per unit volume of the particles. It is not difficult to get the expression for $dQ_{12}/dt$ for particles of different radii, too.

## 4. Numerical calculations of radiative conductance

By definition, the thermal conductance is given by

$$G(z_0) = \dot{Q}(z_0)/\Delta T \quad (16)$$

where $\dot{Q}(z_0)$ is the heating (cooling) rate of one of the interacting bodies having temperatures $T_1$ and $T_2$ ( $\Delta T = T_1 - T_2$ ), at the definite gap width $z_0$. The function $G(z_0)$ may be calculated both at finite $\Delta T$ and in the limit $\Delta T \to 0$. As $\dot{Q}(z_0)$ increases with increasing $\Delta T$, $G(z_0)$ being calculated at $\Delta T \to 0$ turns out to be slightly smaller than assuming $\Delta T$ to be finite.

In the case of normal non magnetic metals, the dielectric functions of the particle and plate have been taken within the Drude model approach with parameters of gold ($\omega_p = 1.37 \cdot 10^{16} \, rad/s, \tau = 1.89 \cdot 10^{-14} \, s$). To analyse the case of insulators, we have chosen silica, which has two infrared absorption domains in the ranges of $0.05 \div 0.07 \, eV$ and $0.14 \div 0.16 \, eV$ [34]. The involved dielectric function has been approximated by simple oscillator model with account of two absorption peaks. The polarizabilities of the particles were taken in classical approximation [32]

$$\alpha_e(\omega) = R^3 \frac{\varepsilon(\omega)-1}{\varepsilon(\omega)+2} \qquad (17)$$

$$\alpha_m(\omega) = -0.5R^3 \left[1 - \frac{3}{R^2 k^2} + \frac{3}{Rk} ctg(Rk)\right], k = (1+\mathrm{i})/\delta(\omega) \qquad (18)$$

where $\varepsilon(\omega)$ is the dielectric permittivity, $\delta(\omega)$ is the skin depth. As we have shown in [33], in calculations of RHE and Casimir forces for metal particles and metal substrates, the contribution of electric polarization proves to be negligibly small, while for dielectric materials like silica one can neglect magnetic polarization effect.

It is important to note that in DAA, when using eq.(18), the particle radius in the square brackets is considered as an additional parameter $\tilde{R}$, different from the particle radius $R$. A choice of $\tilde{R}$ is somewhat arbitrary one. In our work, we have chosen $\tilde{R}$ in so manner that at a minimal gap width of $1 \div 10\, nm$ the resultant thermal conductance proved to be in agreement with that calculated in PFA. So, we have estimated $\tilde{R} \approx 50 \div 60\, nm$ when $R$ has changed in the range of $0.1 \div 50\, \mu m$. The obtained values of $\tilde{R}$ are very close to the characteristic thickness of the skin layer for gold in the case of thermal radiation and, at the same time $\tilde{R} \ll R$ – a necessary condition which must be fulfilled when using DAA.

If use is made of eqs.(8) and (12), one has to integrate quickly oscillating functions related with contributions of propagating waves (due to oscillating denominators in (8)). To calculate the involved integrals, we have developed a numerical method analogous to that one used in [5], representing integrals over wave vectors by sums of integrals over the oscillation periods of the incoming functions.

Figs.2,3 show the calculated thermal conductances of Au particles with radii 5 and $25\,\mu m$ placed near a gold surface $(T = 300 K, \Delta T \to 0)$. The solid and dotted curves correspond to DAA and PFA models. Dashed lines show the contributions of radiation into vacuum (vacuum conductances). For particles with radius $R > 5\,\mu m$, the vacuum conductances have been calculated according to eqs.(9), (16), while for smaller particles with radius $R < 5\,\mu m$ – using the first terms of eqs.(1) and (10).

To compare RHE in configurations sphere –plate and sphere –sphere, Fig.4 shows fractions $K(z_0)$ of the proper radiative conductances (with account of vacuum term) for particles with radii 1,5 and $25\,\mu m$. As can be seen from Fig.4, the assumption [25] that $K(z_0) \approx 2$ is, in general, erroneous. At $z_0 < 0.05\,\mu m$, for particles of radius $R = 5 \div 25\,\mu m$ one gets $K(z_0) > 2$, whereas for smaller particles ($R < 1\,\mu m$) the intensity of thermal radiation in configuration of two spheres may exceed that one in configuration sphere –plate, at a minimal gap width. At $z_0 < 0.1\,\mu m$, as could be expected, both DAA and PFA results are in good agreement. A value of $z_0 = 0.1\,\mu m$ is close to the wave length of surface plasmons and the characteristic length corresponding to fast decrease of radiative heat flux to the vacuum level. However, for distances in the range of $0.1 \div 10\,\mu m$ DAA and PFA models reveal diverse behavior: PFA always gives lower conductance in configuration sphere –plate as compared to the configuration sphere – vacuum, while in DAA the radiative conductance reachs a minimum value at $z_0 \approx 1\,\mu m$ and grows up to the vacuum level with increasing $z_0$, showing small damped oscillations.

Now we touch upon experimental results [25] for the thermal contact $Au-Au$, where the probing tip radius of the scanning thermal microscope was equal to $60nm$, and the tip and plate temperatures were 300K and 100K, respectively. Under these conditions, at a minimum approach distance $z_0 = 1nm$ the measured tip cooling rate was equal to $\dot{Q} \approx 10^{-5}W$. On the other hand, using DAA we obtain $\dot{Q} = 3 \cdot 10^{-9}W$ (at $T_1 = 300K, T_2 = 100K$) and $G = 0.02\,nW/K$ (at $T_1 \to 300K, T_2 = 300K$). One can note an obvious disagreement between the experimental and theoretical estimations. The authors [25] have also noted this fact, using a simplified theoretical model. To overcome the contradiction, they have proposed a special heuristic model. In our opinion, the observed large difference between the theory and experiment is not necessarily related with drawbacks of theoretical models based on fluctuation electromagnetic theory. It may be likely caused by experimental uncertainties of the calibration procedure, which might lead to an overestimation of the measured heat flux. This assumption is favored by good qualitative agreement between the theoretical dependence $\dot{Q}(z_0)$ and the measured one in [25]: both are characterized by the same characteristic length $\Delta z_0$ corresponding to fast decrease of radiation power from maximal values to the vacuum ones. In addition, one observes the same relative change of the radiation power by approximately two orders of value when $z_0$ grows up from $1nm$ to $100nm$. One extra argument in favor of what has been said is that the absolute values of radiation conductance being reported in [25] proved to be much greater than those calculated and measured in the case of dielectric materials (see in what follows).

The calculated thermal conductances for silica particles above glass plates are represented on Figs.5-7. The solid and dashed lines correspond to DAA and PFA models. Figs.5,6 also demonstrate experimental data [26-28] for particles with radii 25 and $47\,\mu m$. In this cases, the plate temperatures were assumed to be 300K, while the temperatures of the particles were assumed to be 346 and $316K$, respectively. As one can see, DAA and PFA models result in several times more intensive thermal radiation at $z_0 < 1\,\mu m$. In the range of $0.3 \div 2\,\mu m$, for particles of radius $25\,\mu m$, DAA results turn out to be in better agreement with experiment, than PFA ones, while for particles of radius $47\,\mu m$ we have an opposite situation. At $z_0 > 1\mu m$, PFA line goes even below the experimental points (cf. Fig.5 and Fig.6). It seems to be probable that the data normalization procedure used in [28] to extract contribution of vacuum conductance, might have some inaccuracy.

As can be seen from Figs. 5,6, at $z_0 < 1\mu m$ radiative conductance of dielectric particles proves to be larger by more than one order of value as compared to that one of metal particles (see Figs.2,3). Moreover, the characteristic decay length for radiative conductance to decrease down to vacuum value increases up to several $\mu m$. This is conditioned by an ability of silica to absorb electromagnetic waves of length 7 and $20\,\mu m$. Therefore, the corresponding near field domain extends to larger distances from the surface of the plates. A comparison between the vacuum radiative conductances of dielectric and metal particles shows the same proportion. Thus, for silica particles with $R = 25\,\mu m$ the vacuum conductance equals $36.6\,nW/K$ (according to eq.(9) at $T = 300K$), whereas for metal particles –only $0.48\,nW/K$. For very small particles with radius $R = 0.05\,\mu m$, according to the first term of eq.(10), the corresponding radiative conductances of silica and gold particles are equal to $0.15\,pW/K$ and $0.002\,pW/K$, i.e. their fraction is nearly the same.

Fig.7 compares thermal conductances in configurations sphere –plate and sphere –sphere. The results show that the fraction $K(z_0)$ of these quantities depends on particle radius $R$ quite differently, as compared to the case of metal particles near a metal surface. Unlike Fig.4, for dielectric particles of $R \leq 1\mu m$ at small gap width we get $K(z_0) > 1$. For bigger particles $K(z_0)$ sharply decreases and heat emitted in configuration of two spheres becomes much greater than in configuration sphere –plate (cf. the lines corresponding to $R = 10\,\mu m$ and $R = 25\,\mu m$). In metal –metal contacts (Fig.4) heat emission of bigger particles in configuration sphere –plate proves to be larger than in configuration of two spheres.

## 5.Conclusions

Based on fluctuation electromagnetic theory, several theoretical models of radiative heat exchange are discussed, allowing to get analytical solutions: configuration of two thick parallel plates, two dipole particles and a dipole particle near a flat surface. When passing to the configuration sphere –plate with small gap width, two approximations are analysed: an analog of the proximity force approximation being used to calculating the van –der -Waals and Casimir

forces –PFA, and a new variant of an additive dipole approximation –DAA. The last approximation is also employed to calculating heat emission in a system of two spherical particles. The heat emission into vacuum space is calculated via SB approximation for particles of large radius, and using the dipole approximation for particles of small radius.

The numerical calculations show that PFA and DAA models provide a rather good agreement with experimental data in silica –glass vacuum contact in the distance range of about $0.2 \div 3 \mu m$. At $z_0 < 0.2 \mu m$ the calculated radiative conductances prove to have greater values. A detailed study of the heat emission in the domain $3 \div 10 \mu m$ necessitates making use a new additional measurements with account of the thermal state of vacuum background.

It is shown that radiative conductances in a vacuum contact of metal bodies turn out to be several dozens times smaller than in contact of dielectric bodies. The characteristic decay length of heat emission for metals is close to $0.1 \mu m$, while for dielectrics like silica it is about several $\mu m$. Some new interesting features are observed: at small gap width, in a vacuum contact between two large dielectric spheres $(R > 5 \mu m)$ radiative conductance is greater than in configuration sphere –plate, and vice versa in contact of metal spheres.

Possible reasons for sharp disagreement between the theoretical estimations and experimental measurements of radiative conductance in vacuum contact between an $Au$ probing tip of the scanning probe microscope and the surface of gold are discussed.

**Acknowledgements**

**Appendix**

$$Y_1(x,y) = \int_1^{1+y} \exp(-xz)\left[y(z-1)-(z-1)^2\right]dz = \frac{(xy+2)}{x^3}\exp(-x(1+y)) + \frac{(xy-2)}{x^3}\exp(-x) \quad (A1)$$

$$Y_2(x,y) = \int_1^{1+y} \cos(xz)\left[y(z-1)-(z-1)^2\right]dz = -\frac{y}{x^2}\cos(x(1+y)) +$$
$$+ \frac{2}{x^3}\sin(x(1+y)) - \frac{y}{x^2}\cos x - \frac{2}{x^3}\sin x \quad (A2)$$

$$Y_3(x,y) = \int_1^{1+y} \sin(xz)\left[y(z-1)-(z-1)^2\right]dz = -\frac{y}{x^2}\sin(x(1+y)) -$$
$$- \frac{2}{x^3}\cos(x(1+y)) - \frac{y}{x^2}\sin x + \frac{2}{x^3}\cos x \quad (A3)$$

$$f_1(x) = \frac{3}{32}\left[\ln(1-x^2) + \frac{x^2(2-x^2)}{2(1-x^2)}\right] \quad (A4)$$

$$f_2(x) = \frac{3}{32}\left[2x^2 - (3x^2 - 2)\ln(1-x^2) - x^3 \ln\frac{1+x}{1-x}\right] \quad (A5)$$

$$f_3(x) = \frac{3}{320}\left[11x^4 + 2x^2 - (10x^3 - 2x^5)\ln\frac{1+x}{1-x} - (10x^2 - 2)\ln(1-x^2)\right] \quad (A6)$$

**Captions to the figures**

Fig.1 Plotted in this figure are separate contributions to the radiative conductance of small metal particles (in relative units) computed from eq.(1): 1 –evanescent modes, 2 – surface wave modes, 3 –vacuum modes (the first term (1)), 4 – resulting dependence on $z_0$.

Fig.2 Radiative conductance of spherical $Au$ particles in vacuum contact with a gold surface ($R = 5\mu m, T = 300K, \Delta T \to 0$). Solid line –DAA with account of vacuum radiation, dotted line – PFA, dashed –dotted line –the contribution of vacuum radiation according to the first term of eq. (1).

Fig.3. The same as on Fig.2 at $R = 25\mu m$. The dashed –dotted line was computed via SB approximation –eq.(9) and describes vacuum radiation.

Fig.4 Plotted is the fraction of radiative conductances in configurations sphere –plate (DAA) and sphere -sphere (eq.(10)) for $Au$ particles near a gold surface ($T = 300K, \Delta T \to 0$). Vacuum radiation is taken into account. The particles radii expressed in $\mu m$ are indicated.

Fig.5 Radiative conductance of spherical silica particles with radius $R = 47\mu m$ in vacuum contact with a glass plate: $T_1 = 316K, T_2 = 300K$. Solid line –DAA (without vacuum radiation), dotted line –PFA, circles –experimental data [26,27]. The results of DAA calculation were additionally increased by $0.1 nW/K$ in order to get positive values of radiative conductance in overall range of distances.

Fig.6 The same as on Fig.5 at $R = 25\mu m$, $T_1 = 346.5K, T_2 = 300K$. The experimental data [28] are shown by circles.

Fig.7 The same as on Fig.4 for silica particles near a glass plate.

Fig.1

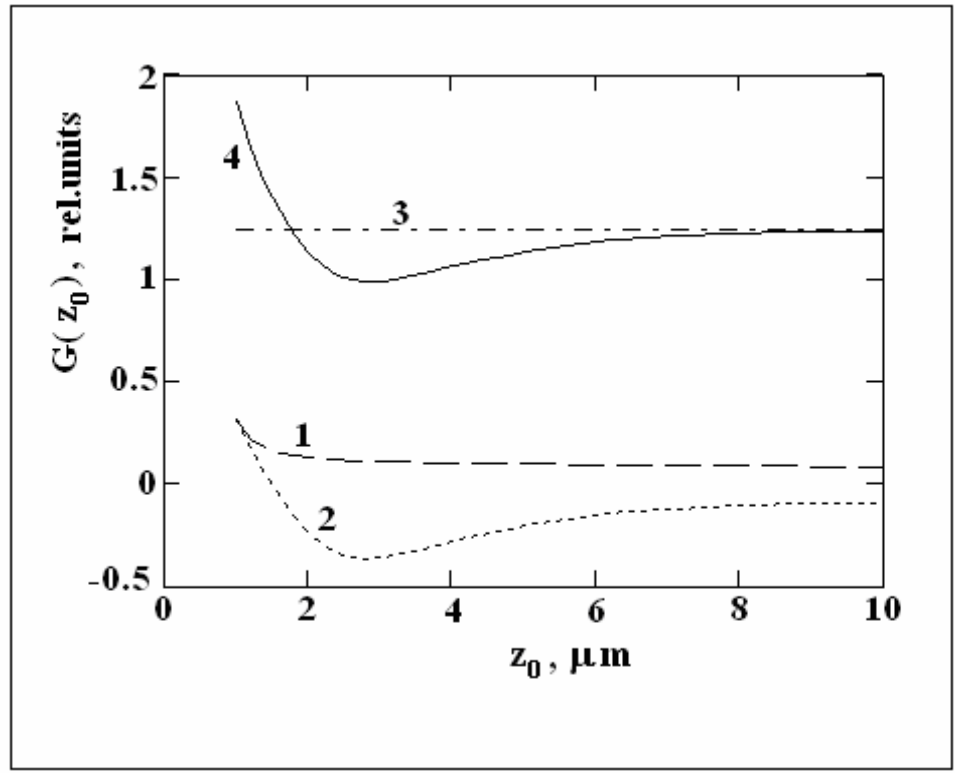

Fig.2

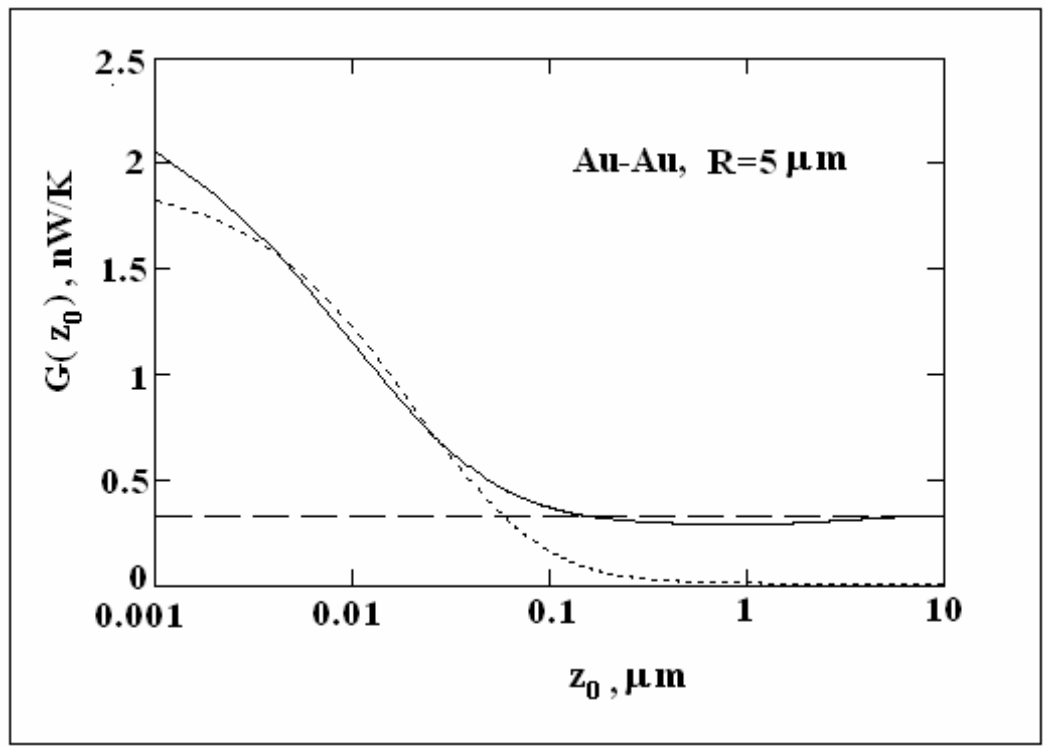

Fig.3

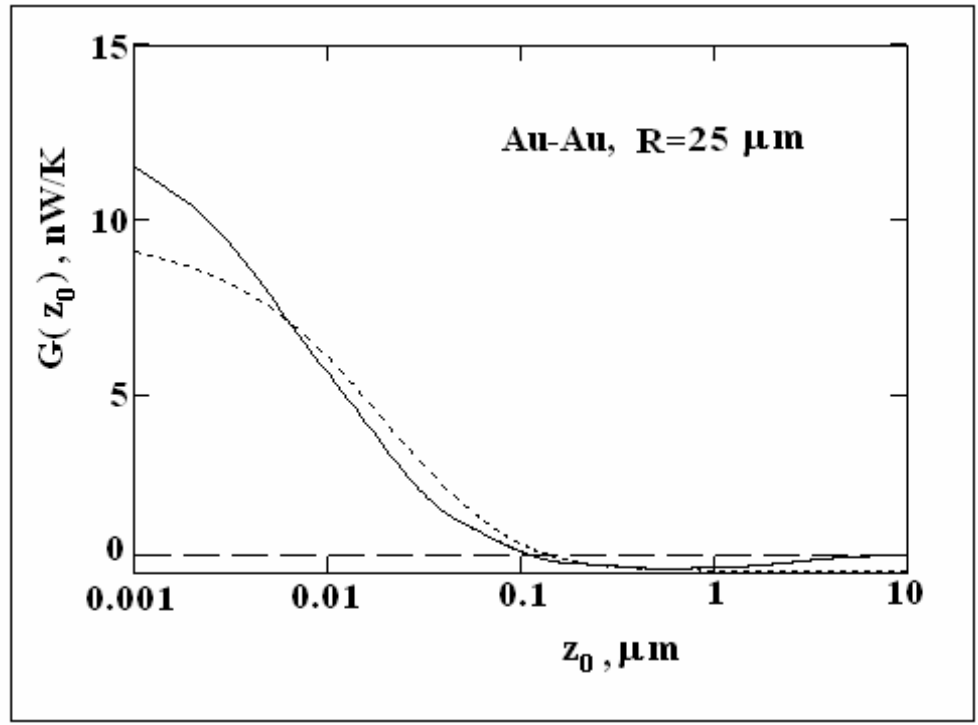

Fig.4

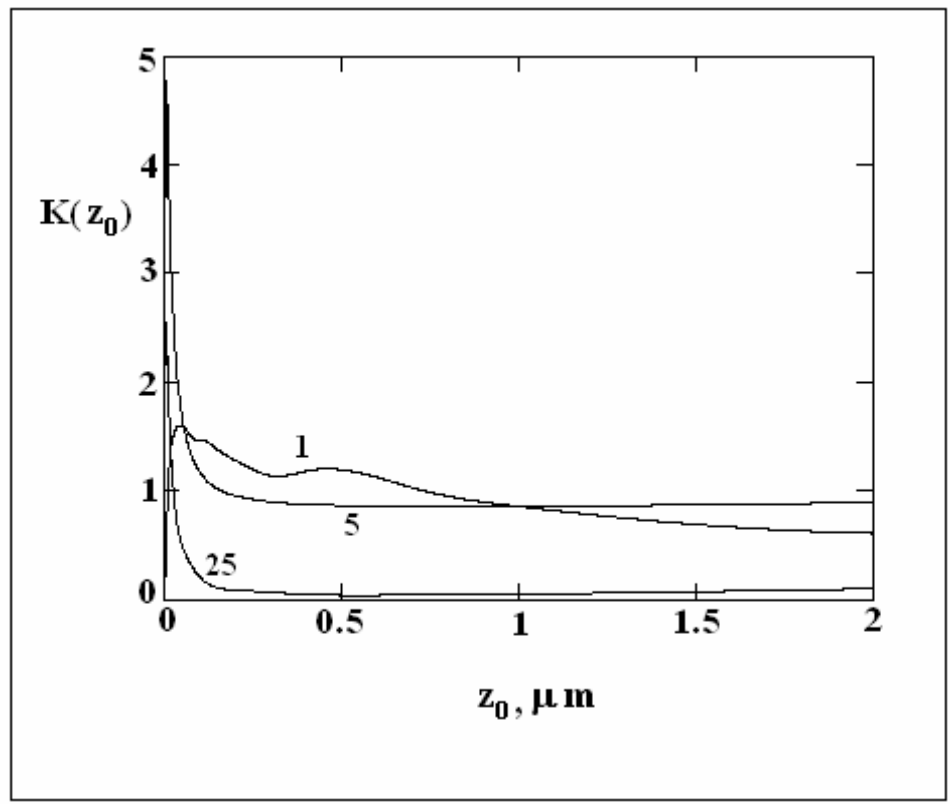

Fig.5

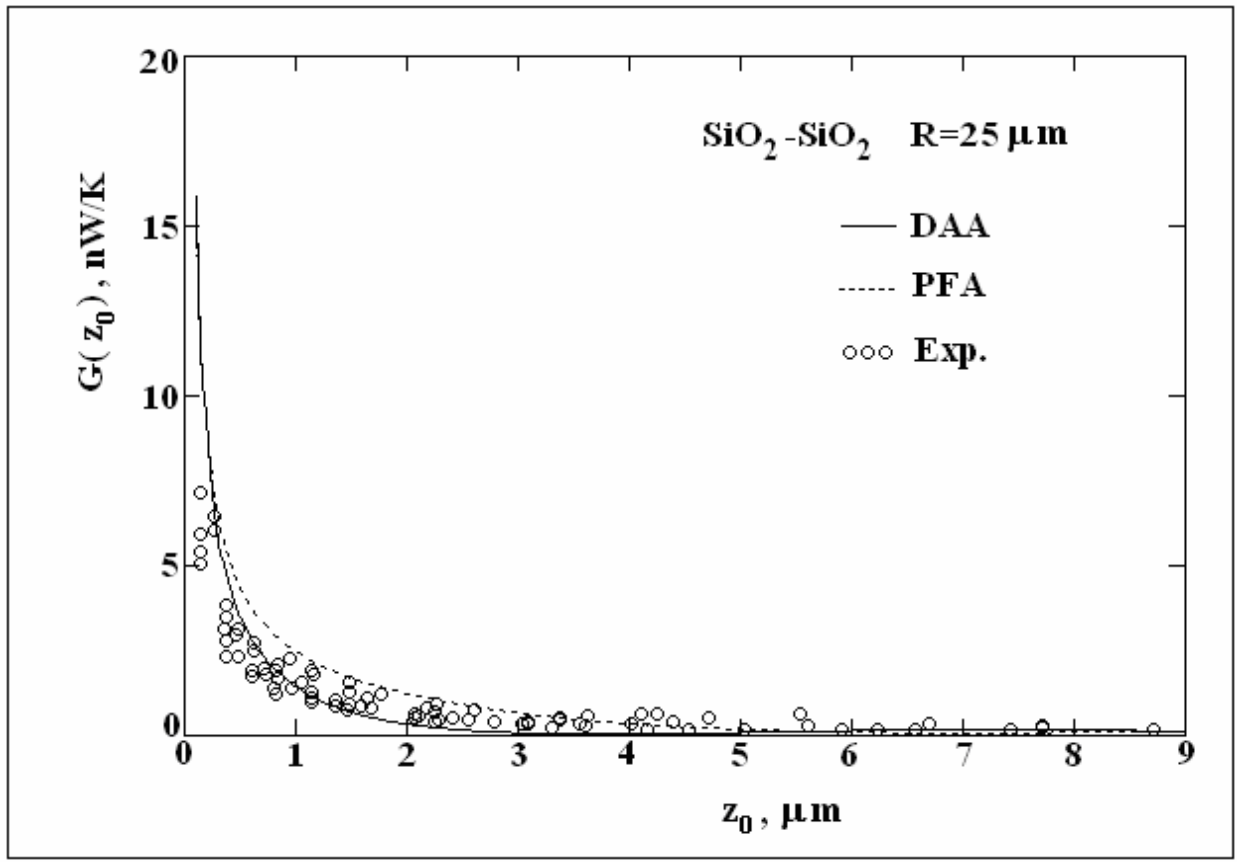

Fig.6

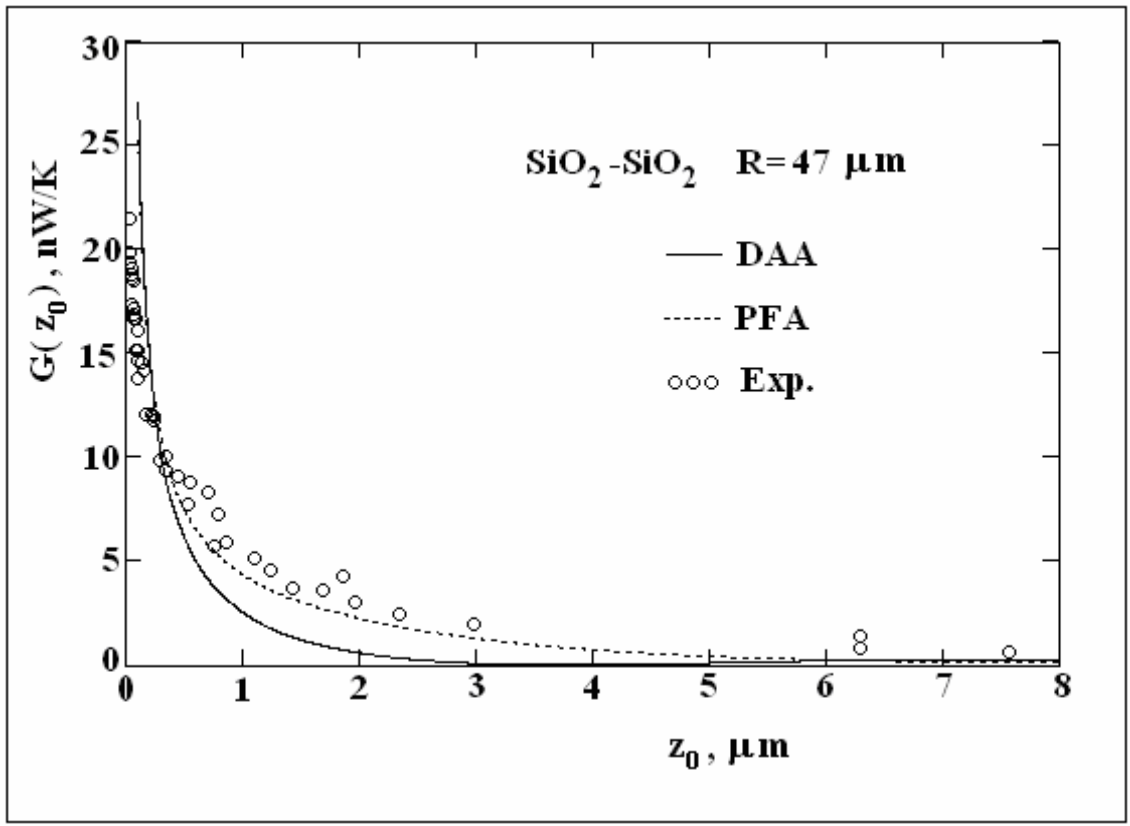

Fig.7

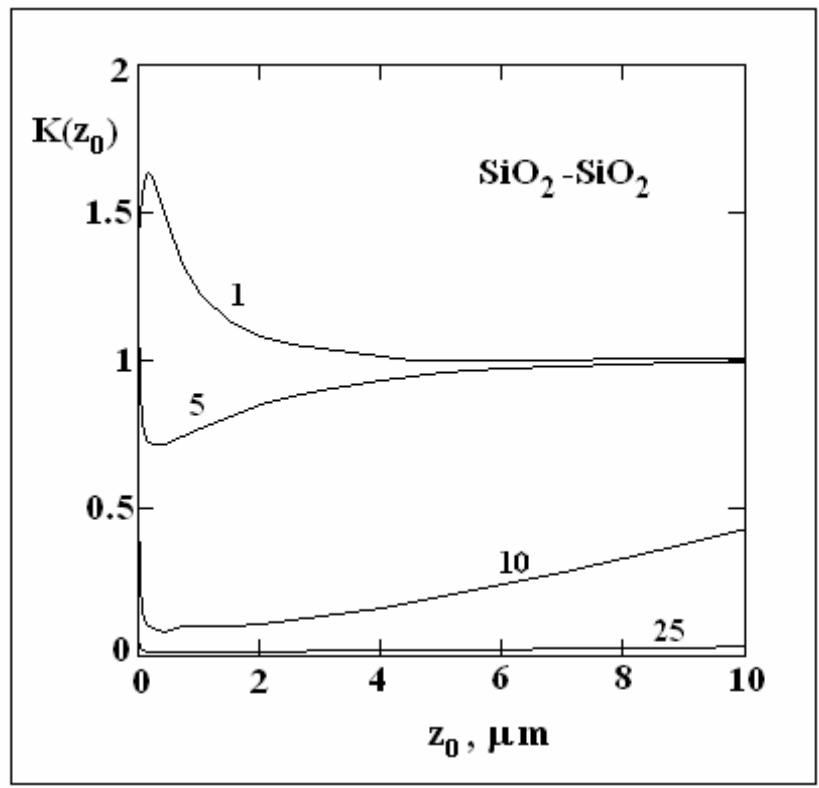